\documentclass[conference]{IEEEtran}
\IEEEoverridecommandlockouts
\usepackage{cite}
\usepackage{amsmath,amssymb,amsfonts}
\usepackage{algorithmic}
\usepackage{graphicx}
\usepackage{textcomp}
\usepackage{xcolor}
\usepackage{hyperref}
\usepackage{makecell}
\usepackage{booktabs}

\def\BibTeX{{\rm B\kern-.05em{\sc i\kern-.025em b}\kern-.08em
    T\kern-.1667em\lower.7ex\hbox{E}\kern-.125emX}}
\begin{document}

\title{Adaptive Hierarchical Evaluation of LLMs and SAST tools for CWE Prediction in Python
\thanks{This work is funded under the agreement with the ACT Government, Future Jobs Fund - Open Source Institute(OpenSI) - R01553; and NetApp Technology Alliance Agreement with OpenSI - R01657.
Additionally, this research was supported by the Australian Government through the Department of Education’s National Industry PhD Program (project 36337). 
The views expressed herein are those of the authors and are not necessarily those of the Australian Government or the Department of Education.}
}

\author{\IEEEauthorblockN{Muntasir Adnan}
\IEEEauthorblockA{\textit{Open Source Institute} \\
\textit{Faculty of Science and Technology}\\
\textit{University of Canberra}\\
Canberra, Australia \\
Adnan.adnan@canberra.edu.au}
*Corresponding author
~\\
\and
\IEEEauthorblockN{Carlos C. N. Kuhn}
\IEEEauthorblockA{\textit{Open Source Institute} \\
\textit{Faculty of Science and Technology}\\
\textit{University of Canberra}\\
Canberra, Australia \\
Carlos.NoschangKuhn@canberra.edu.au}
}

\maketitle

\begin{abstract}
Large Language Models have become integral to software development, yet they frequently generate vulnerable code. 
Existing code vulnerability detection benchmarks employ binary classification, lacking the CWE-level specificity required for actionable feedback in iterative correction systems. 
We present ALPHA (Adaptive Learning via Penalty in Hierarchical Assessment), the first function-level Python benchmark that evaluates both LLMs and SAST tools using hierarchically aware, CWE-specific penalties. 
ALPHA distinguishes between over-generalisation, over-specification, and lateral errors, reflecting practical differences in diagnostic utility. 
Evaluating seven LLMs and two SAST tools, we find LLMs substantially outperform SAST,
though SAST demonstrates higher precision when detections occur. 
Critically, prediction consistency varies dramatically across models (8.26\%-81.87\% agreement), with significant implications for feedback-driven systems. 
We further outline a pathway for future work incorporating ALPHA penalties into supervised fine-tuning, which could provide principled hierarchy-aware vulnerability detection pending empirical validation.
\end{abstract}


\section{Introduction}
Large Language Models (LLMs) have fundamentally transformed software development practices, with code-generation capabilities now integrated into mainstream development workflows via tools such as GitHub Copilot, Amazon CodeWhisperer, and ChatGPT.
This paradigm shift has dramatically improved developer productivity, enabling rapid prototyping and accelerating development cycles~\cite{soft_eng_game_changer}.
However, recent studies demonstrate that LLMs frequently produce code containing exploitable vulnerabilities, with security flaws appearing in 40-60\% of generated code snippets, depending on task complexity and model~\cite {asleep, how_secure}.
The research community has approached this vulnerability challenge through two primary avenues: prompt engineering and supervised fine-tuning.

Prompt engineering techniques have demonstrated potential, with security-focused prompts reducing vulnerability generation by up to 61\% in models like GPT-4o~\cite{prompt_gpt}.
Self-reflection approaches, where models detect vulnerability in their previously generated code and apply repairs, achieve 41.9-68.7\% vulnerability remediation when specifically prompted~\cite{prompt_gpt, llm_correct}.
However, these techniques exhibit significant practical limitations.
Not only do they require vulnerability-specific instructions to achieve substantial improvements, but more critically, they demonstrate concerning instability.
Results vary substantially with minor prompt modifications and prove inconsistent both within and across studies~\cite{role_play, neg_role_play}. 

Fine-tuning approaches have demonstrated improvements~\cite{case_study, instruction_tuning, sven}, but suffer from critical limitations including catastrophic forgetting~\cite{cat_forgetting}, poor generalisation~\cite{guiding_ai} and prohibitive computational costs for frequent retraining as vulnerability landscapes evolve~\cite{cat_forgetting, reversal_curse}.
Recent evidence suggests that contemporary Small Language Models (SLMs) with improved training regimes can match or exceed the performance of older, larger models~\cite {nvidia_slm}, potentially rendering vulnerability-specific fine-tuning obsolete.

Given these limitations, iterative feedback loops have emerged as a promising alternative, already demonstrating effectiveness in improving functional correctness in code generation~\cite{mapcoder, pycapsule, ddi}.
These systems iteratively analyse generated code, identify specific issues, and prompt the model to address them.
However, a critical question remains unanswered: what tool should provide the feedback for vulnerability detection?
Two candidate approaches exist: traditional static analysis security testing (SAST) tools and LLM-based vulnerability detection.
Current practice predominantly employs SAST tools for feedback generation~\cite{guiding_ai, static_feedback}, yet this choice appears conventional rather than evidence-based.
Recent comparative studies suggest LLMs can outperform SAST tools for vulnerability detection~\cite{castle, static_vs_llm}, though results remain limited, mixed and context-dependent.

\subsection{Research Gap and Contribution}
Existing SAST-LLM comparisons suffer from a fundamental methodological limitation: most employ binary classification (vulnerable/not vulnerable)~\cite{pyvul, static_vs_llm}, which provides insufficient specificity for iterative correction.
Effective feedback mechanisms require precise vulnerability identification, specifically the CWE type, location, and nature of the flaw~\cite{llm_correct}.
Without this granularity, models can not receive the actionable, vulnerability-specific instructions that prompt-engineering studies suggest are necessary for substantial security improvements~\cite{prompt_gpt}.

Whilst some studies compare SAST and LLMs~\cite{static_vs_llm, pyvul}, none provide the benchmark needed to inform iterative feedback loop design at the function level for Python.
This specificity is essential; function-level analysis respects LLM context window constraints and enables granular debugging~\cite{pycapsule}.
Python represents the most critical use case given its ubiquity in development workflows, where LLM code generation is most prevalent.
This work addresses this gap directly.

We present the first comprehensive benchmark that compares multiple LLMs (including modern SLMs) and SAST tools across multiple Python vulnerability datasets, using CWE-level granularity at the function level.
Our contributions are:
\begin{itemize}
    \item The first function-level Python vulnerability detection benchmark - \textbf{ALPHA} (Adaptive Learning via Penalty in Hierarchical Assessment), evaluating both LLMs and SAST tools on CWE-specific identification, providing an empirical foundation for feedback mechanism selection.
    
    \item Analysis of detection consistency across multiple runs for each LLM, assessing reliability for iterative correction systems where consistent feedback is paramount.
    
\end{itemize}

\section{Related Work}
Early efforts to improve code security focused on fine-tuning LLMs to generate secure code~\cite{sven, instruction_tuning}. 
Studies employed vulnerability token classification~\cite{instruction_tuning} and adversarial training~\cite{sven}, reporting 30-32\% accuracy improvements. 
However, finetuned models tend to exhibit poor generalisation and overfitting~\cite{guiding_ai}, with documented instances of the reversal curse where fine-tuned models lose knowledge symmetry~\cite{reversal_curse}.
Catastrophic forgetting~\cite{cat_forgetting} causes abrupt loss of previously learned tasks, while the substantial computational cost of fine-tuning~\cite{peft, unleashing} renders the frequent retraining needed to track newly emerging vulnerabilities economically infeasible.
Critically, a vulnerability augmented prompting framework DLAP~\cite{dlap} found prompt engineering superior to fine-tuning for both vulnerability detection accuracy and cost-effectiveness, leading us to examine prompting-based approaches.

Prompt engineering offers a computationally efficient alternative, with reflection-based approaches showing particular promise. 
A recent study~\cite{prompt_gpt} using GPT models reported a 61\% improvement with recursive criticism and improvement (RCI), in which models iteratively critique and refine their outputs.
However, prompt-based approaches demonstrate concerning instability. 
Persona-based prompting yielded highly inconsistent results across studies. 
Kong and Zhao~\cite{role_play} reported 60\% accuracy gains, Bruni et al.~\cite{prompt_gpt} reported 43\% improvement with GPT models, whilst Zheng et al.~\cite{neg_role_play} found persona prompts had random or negative effects across four LLM families. 
This sensitivity even extends to variations in wording~\cite {prompt_gpt}, undermining reliability.

Given the impracticality of fine-tuning and the instability of prompt engineering, iterative feedback loops have emerged as the most promising approach, demonstrating 15-35\% improvement in Python code generation~\cite{mapcoder, pycapsule, ddi}.
Feedback loops combine the efficiency of prompt engineering with structured external validation and feedback.
Which raises the critical question: which tool should generate feedback for vulnerability repair?  
Notably, all existing implementations of iterative feedback loop for code vulnerability detection rely on SAST tools and use LLMs primarily for interpretation and code repair~\cite{static_feedback, patch_sec, guiding_ai, securefixagent}.
Furthermore, SAST tools frequently serve as ground truth for evaluation datasets~\cite{guiding_ai, securefixagent}, creating a methodological circularity in which the appropriateness of these tools for vulnerability detection remains unexamined.
Thus, the preference for SAST tools in feedback loops appears empirically driven rather than systematically validated. 

Direct comparisons between SAST tools and LLMs for vulnerability detection yield mixed and context-dependent results. 
For C programs in smaller codebases, Castle~\cite{castle} reported LLM superiority (GPT-3.5-mini: 977/1250 vs. best static tool: 661/1250), whilst Gong et al.~\cite{end_to_end_code} found GPT-4 achieved 76\% detection with only 3.1\% false positives yet still recommended static tools over LLMs. 
Python-focused studies show similarly varied results: Li et al.~\cite{everything_wanted} achieved 67\% accuracy with DeepSeek R1, Zhou et al.~\cite{static_vs_llm} reported LLM scores reaching 100 versus SAST's 44.4 at repository-level analysis, whilst PyVul~\cite{pyvul} found CodeQL managed only 10.8\% accuracy at commit-level vulnerability detection compared to LLMs' 49.5-58.5\% at function-level.

Most studies perform binary classification (vulnerable/not vulnerable)~\cite{everything_wanted, static_vs_llm, pyvul} rather than CWE-level identification, evaluate at the repository or commit level rather than the function level, and, critically, none assess detection stability across multiple runs. 
For iterative feedback systems, these limitations are fatal: binary classification cannot guide targeted remediation, repository-level analysis exceeds LLM context windows, and unstable feedback undermines iterative improvement.

No existing study directly compares SAST and LLMs as feedback tools for detecting function-level Python vulnerabilities at CWE-level granularity — the precise requirements for practical iterative feedback systems.
Given Python's dominance in LLM code generation and context window constraints that necessitate function-level analysis, determining which tool provides more reliable, actionable feedback remains an open question with direct practical implications.

\section{Methodology}

\subsection{Datasets}
\label{subsec: dataset}
We evaluate on two Python vulnerability datasets with distinct characteristics.
Dataset selection was constrained by three requirements: function-level granularity, explicit CWE ground truth labels, and human-annotated labels rather than SAST-derived ground truth.
SecurityEval~\cite{siddiq2022seceval} contains 121 vulnerable functions distributed across the CWE hierarchy: 73 base weaknesses, 26 class weaknesses, 16 variant weaknesses, and 3 pillar weaknesses. 
SVEN~\cite{sven} comprises 342 Python samples (filtered from the original 720 multi-language dataset), all labelled as base weaknesses. 
This difference in the distribution of abstraction levels is significant: SecurityEval's hierarchical diversity creates a more challenging detection task with varying specificity requirements, whereas SVEN's uniform base-level labels provide more precise prediction targets. 
Both datasets provide ground-truth CWE labels suitable for evaluating hierarchically-aware detection.
\subsection{CWE Graph Construction}
CWE provides a hierarchical taxonomy of software weaknesses.
We construct a direct graph $G = (V, E)$ from the MITRE CWE database, where - 
\begin{itemize}
    \item Nodes ($V$): CWE weakness types specifically: \textit{pillar weakness, class weakness, base weakness, variant weakness, compound weakness} and \textit{chain} weakness.
    \item Edges ($E$): Parent-child relationship from the CWE hierarchy.
\end{itemize}
We excluded two CWE types—\textit{view} and \textit{category}, which function as organisational constructs in the CWE database.  
This removes 49 nodes and 480 edges in total without losing any generalisation. 
\textit{Views} present alternative perspectives on weaknesses (e.g., by development phase or audience), while \textit{Categories} group weaknesses by shared characteristics (e.g., authentication errors). 
Unlike weakness types (\textit{pillar, class, base, variant}), these constructs do not describe specific, detectable vulnerabilities but serve as navigational aids within the taxonomy.
The resulting graph contains 944 nodes and 1153 edges as opposed to 993 nodes and 1633 edges in the full graph with all CWES.

\subsection{ALPHA: Adaptive Penalty Metric}
For a given code sample with ground truth CWE $c_{true}$ and predicted CWE $c_{pred}$, we define the
penalty score $P$:
\begin{equation*}
    P(c_{\text{pred}}, c_{\text{true}}) = d(c_{\text{pred}}, c_{\text{true}}) \times \alpha(c_{\text{pred}}, c_{\text{true}})
\end{equation*}
where the distance component $d(c_{pred}, c_{true})$ is the shortest path length between nodes in the graph $G$, computed using Dijkstra's algorithm\cite{dijk} on the undirected representation. 
This captures how ``far" the prediction deviates from $c_{true}$ in the weakness taxonomy.
The direction multiplier $\alpha(c_{pred}, c_{true})$ encodes the semantic type of error - 
\[
\alpha(c_{\text{pred}}, c_{\text{true}}) =
\left\{
\begin{array}{@{}l l@{}}
\alpha_{\text{up}}
& \makecell[l]{\text{if } c_{\text{pred}} \text{ is an ancestor of } c_{\text{true}}\\
(generalising)} \\[4pt]
\alpha_{\text{down}}
& \makecell[l]{\text{if } c_{\text{pred}} \text{ is a descendant of } c_{\text{true}}\\
(over-specifying)} \\[4pt]
\alpha_{\text{lateral}}
& \makecell[l]{\text{otherwise (lateral error)}}
\end{array}
\right.
\]
Direction determination uses the directed graph, where $c_{pred}$ is an ancestor of $c_{true}$ if there exists a directed path from $c_{pred}$ to $c_{true}$, descendant if the reverse holds; lateral otherwise.
While the shortest path distance captures error magnitude, it treats all errors at the same distance as equally severe.
This fails to capture important asymmetries in practical utility.
We distinguish three error types:
\begin{itemize}
    \item \textbf{Generalising Errors ($\alpha_{up} = 2.0$)}: When predictions move up the hierarchy, critical diagnostic information is lost. 
    For example, predicting CWE-707 (Improper Neutralisation, Pillar) when the true weakness is CWE-89 (SQL Injection, Base) provides minimal actionable guidance. 
    We assign the highest fixed penalty of 2.0 to reflect this severe loss of utility.

    \item \textbf{Lateral Errors ($\alpha_{lateral} = 1.8$)}: Predictions in different branches indicate conceptual misunderstanding but may retain some diagnostic value if weaknesses share parent classes or belong to related domains. 
    We assign a moderate fixed penalty of 1.8, lower than generalisation but higher than over-specification, reflecting partial correctness in the general weakness category.

    \item \textbf{Over-Specifying Errors ($\alpha_{down},$ Adaptive)}: When predictions move down the hierarchy, errors introduce unjustified specificity while maintaining correct general understanding.
    Crucially, severity depends on available ``room for specificity" in the ground truth's descendant hierarchy. 
    We define:
    \begin{equation*}
        \alpha_{down}(c_{true}) = \alpha_{max} - (\alpha_{max} - \alpha_{min}) \times \frac{\delta(c_{true})}{\Delta(type(c_{true}))}
    \end{equation*}
    where $\delta(c_{true})$ is the maximum distance from $c_{true}$ to any descendant leaf in the family subtree (depth), $\Delta(type(c_{true}))$ represents the maximum $\delta$ among all CWEs of that type.
    Here $\alpha_{max} = \alpha_{lateral} = 1.8$ and $\alpha_{min} = 1.1$.

    The rationale for such adaptive approach is - 
    \begin{itemize}
        \item Leaf nodes ($\delta = 0$): When $c_{true}$ is a leaf with no children, predicting something more specific is semantically equivalent to a lateral error.
        The prediction makes unjustified taxonomic assumptions about specificity that doesn't exist. 
        Thus, $\alpha_{max} = \alpha_{lateral} = 1.8$
    
        \item Deep subtrees (high $\delta$): When $c_{true}$ is a \textit{Class} or \textit{Pillar} with many descendants, substantial ``room for specificity" exists. 
        Over-specification is a mild error, the prediction is conceptually correct but assumes a specific instantiation. 
        Thus, $\alpha_{min} = 1.1$ (exceeds 1.0 to remain penalty-effective).
    \end{itemize}
\end{itemize}
The over-specification penalty depends on ground truth context; the same distance has different severity depending on whether $c_{true}$ permits further refinement. 
Unlike over-specification, generalisation always loses information regardless of ground truth characteristics.
The severity is uniformly high.
Thus, a fixed high penalty appropriately captures this consistent harm.
The parameter values ($\alpha_{up} = 2.0$, $\alpha_{\text{lateral}} = 1.8$, 
$\alpha_{\text{min}} = 1.1$) maintain the penalty hierarchy $\alpha_{\text{up}} > 
\alpha_{\text{lateral}} > \alpha_{\text{down}}$ whilst ensuring both distance and 
direction contributes meaningfully to final scores.

\subsection{Handling Edge Cases}
An out-of-graph prediction where $c_{pred} \not \in V$ suggests that $c_{pred}$ is likely a \textit{category}, \textit{view}, or invalid CWE.
For such cases, we assign the maximum penalty multiplier $\alpha_{oog} = 2.5$ and a distance based on the graph structure $d_{oog} = \left\lceil \bar{d} \right\rceil$
where $\bar{d}$ is the mean shortest path distance across all connected CWE pairs in $G$:

\begin{equation*}
    \bar{d} = \frac{1}{|C|} \sum_{(u,v) \in C} d_G(u,v)
\end{equation*}
Where $C$ is the set of all connected pairs of CWE nodes in $G$, with $|C| = 445,096$, yielding $\bar{d} = 6.26$, which gives $d_{oog} = \lceil 6.26 \rceil = 7$.
This choice positions out-of-graph errors above the typical in-graph error severity, ensuring that hallucinated or overly vague predictions receive higher base penalties than the average in-graph mistake.
Combined with $\alpha_{oog} = 2.5$, this yields a penalty of $P = 7 \times 2.5 = 17.5$, heavily penalising predictions that fall outside the vulnerability hierarchy.

If no path exists between $c_{pred}$ and $c_{true}$ within the graph (different connected components), we apply the same out-of-graph penalty $(d_{oog}, \alpha_{oog})$, indicating completely unrelated weakness types.

When a CWE has multiple parents via different relationship paths, we use the shortest path for distance calculation. 
Direction is determined by whether any directed path exists in the specified direction.

\subsection{Evaluation Protocol}
For each test sample $i$ with ground truth $c_{true}^i$ and prediction $c_{pred}^i$:
\begin{itemize}
    \item Compute individual penalty 
        \begin{equation*}
            P_i = P(c_{pred}^i, c_{true}^i)
        \end{equation*}
    \item Aggregate across dataset - 
        \begin{equation*}
            ALPHA = \frac{1}{N}\sum_{i = 1}^N P_i
        \end{equation*}
\end{itemize}
A lower average penalty indicates better performance.

\section{Experiment Results}
All LLM experiments were conducted 3 times on an NVIDIA A100-PCIE-40GB GPU.
We evaluated seven open-source models: Qwen2.5-Coder-32B~\cite{qwen25coder}, Qwen2.5-Coder-7B~\cite{qwen25coder}, Mistral-7B~\cite{jiang2023mistral7b}, Llama3.1-8B~\cite{llama}, Phi4-14B~\cite{abdin2024phi}, DeepSeek-Coder-6.7B~\cite{guo2024deepseek}, and Devstral-24B~\cite{devstral}.
The complete prompting strategy and templates are detailed in Appendix~\ref{app: system_prompt}.

The SAST experiments were conducted once, as these tools produce deterministic outputs, unlike stochastic LLM inference.
We evaluated two static analysis tools: CodeQL~\cite{codeql} (v2.23.0) and Semgrep~\cite{semgrep} (v1.136.0).
For CodeQL, we used the Python security-extended and security-experimental query suites, which collectively cover 123 CWEs through 101 built-in queries~\cite{pyvul}. 
CodeQL performs semantic analysis by converting source code into a queryable database while preserving program semantics, including data and control flows. 
For Semgrep, we employed four rulesets: python (general Python rules), security-audit, owasp-top-ten, and bandit (39 pattern-matching rules covering 17 CWEs)~\cite{static_vs_llm}. 
Both tools output results in structured formats (SARIF for CodeQL, JSON for Semgrep) for automated parsing.

All experiments used Ollama default parameters (temperature=0.8) to reflect realistic deployment conditions. 
This introduces stochasticity that we explicitly measure through our consistency analysis in Section~\ref{subsec: con2}

\subsection{Contribution 1}
\begin{table}[t]
\caption{ALPHA scores for vulnerability detection for SecurityEval and SVEN Python. 
Lower scores indicate better performance. 
LLM results show mean $\pm$ standard deviation across 3 runs.}
\label{tab: alpha_results}
\begin{center}
\begin{tabular}{|c|c|c|}
\hline
\textbf{Model} & \textbf{SVEN} & \textbf{SecurityEval} \\
\hline
Qwen2.5-Coder-32B & 574.7 $\pm$ 42.4 & 617.8 $\pm$ 18.5 \\
\hline
Mistral-7B & 923.3 $\pm$ 77.0 & 1156.9 $\pm$ 32.0 \\
\hline
Qwen2.5-Coder-7B & 753.2 $\pm$ 39.5 & 921.5 $\pm$ 15.1 \\
\hline
Llama3.1-8B & 813.5 $\pm$ 51.0 & 856.5 $\pm$ 42.7 \\
\hline
Phi4-14B & 646.4 $\pm$ 47.4 & 797.4 $\pm$ 13.0 \\
\hline
DeepSeek-Coder-6.7B & 2557.0 $\pm$ 194.3 & 1228.5 $\pm$ 60.9 \\
\hline
Devstral-24B & 695.1 $\pm$ 71.2 & 578.5 $\pm$ 27.4 \\
\hline
\multicolumn{3}{|c|}{\textbf{SAST Tools}} \\
\hline
CodeQL (Confidence) & 5981.9 & 1261.0 \\
\hline
CodeQL (Any Match) & 5981.9 & 1212.9 \\
\hline
Semgrep (Confidence) & 3663.5 & 1307.6 \\
\hline
Semgrep (Any Match) & 3562.7 & 1307.6 \\
\hline
\end{tabular}
\end{center}
\end{table}

Table~\ref{tab: alpha_results} presents ALPHA scores for seven LLMs and two SAST tools across the SVEN and SecurityEval datasets, where lower scores indicate better CWE-specific detection performance. 
For LLMs, we report mean and standard deviation across three independent runs to account for stochastic variation.
Model rankings exhibit strong consistency across datasets (Spearman $\rho = 0.86, p=0.014 \text{ for LLMs; } \rho = 0.92, p<0.001$ when including SAST tools), demonstrating that ALPHA provides stable evaluations regardless of the specific vulnerability samples. 
Qwen2.5-Coder-32B and Devstral-24B achieve the lowest penalties on both datasets, while DeepSeek-Coder-6.7B shows notably high penalties throughout.

SAST tools exhibit substantially higher ALPHA scores than most LLMs, primarily due to low detection coverage.
CodeQL and Semgrep produced no predictions for over 50\% of the samples. 
When they do predict, CodeQL demonstrates higher precision (72.4\% of predictions contain the ground truth CWE) compared to Semgrep (50.0\%), highlighting a coverage-precision trade-off.

For SAST tools, we report two ALPHA scores corresponding to different CWE selection strategies, since these tools often report multiple potential vulnerabilities per sample. 
The \textit{confidence} strategy selects the CWE with the highest confidence rating (or first reported); the \textit{any-match} strategy gives credit if the ground truth appears anywhere in the output. 
The minimal difference between these scores indicates that when SAST tools report multiple CWEs, the ground truth is rarely among them.
Their errors stem from missed detections rather than incorrect prioritisation.

\subsection{Contribution 2}
\label{subsec: con2}
\begin{table}[b]
\caption{LLM predictions across three runs for SVEN dataset}
\label{tab: consistency_sven}
\begin{center}
\begin{tabular}{|c|c|c|}
\hline
\textbf{\shortstack{Model}} & \textbf{\shortstack{Perfect\\Agreement (\%)}} & \textbf{\shortstack{Majority\\Agreement (\%)}} \\
\hline
deepseek-coder:6.7b & 23.98 & 72.51 \\
\hline
devstral:24b & 73.39 & 91.81 \\
\hline
llama3.1:8b & 50.88 & 75.73 \\
\hline
mistral:7b & 63.74 & 86.84 \\
\hline
phi4:14b & 66.96 & 91.23 \\
\hline
qwen2.5-coder:32b & 81.87 & 92.98 \\
\hline
qwen2.5-coder:7b & 71.64 & 89.77 \\
\hline
\end{tabular}
\end{center}
\end{table}


\begin{table}[t]
\caption{LLM predictions across three runs for SecurityEval dataset}
\label{tab: consistency_sec_eval}
\begin{center}
\begin{tabular}{|c|c|c|}
\hline
\textbf{Model} & \textbf{\shortstack{Perfect\\Agreement (\%)}} & \textbf{\shortstack{Majority\\Agreement (\%)}} \\
\hline
deepseek-coder:6.7b & 8.26 & 49.59 \\
\hline
devstral:24b & 52.07 & 85.95 \\
\hline
llama3.1:8b & 16.53 & 55.37 \\
\hline
mistral:7b & 20.66 & 56.20 \\
\hline
phi4:14b & 34.71 & 71.90 \\
\hline
qwen2.5-coder:32b & 61.98 & 86.78 \\
\hline
qwen2.5-coder:7b & 21.49 & 62.81 \\
\hline
\end{tabular}
\end{center}
\end{table}

For iterative correction systems, consistent feedback is essential. 
A model producing different vulnerability assessments for identical code cannot provide reliable feedback signals. 
We evaluate detection consistency by running each LLM three times on both datasets under identical conditions, measuring agreement across runs.
Tables~\ref{tab: consistency_sven} and~\ref{tab: consistency_sec_eval} present two agreement metrics: \textit{Perfect Agreement} (all three runs produce identical CWE predictions) and \textit{Majority Agreement} (at least two runs concur). 

Consistency varies dramatically across models. 
Perfect Agreement ranges from 81.87\% (Qwen2.5-Coder-32B on SVEN) to just 8.26\% (DeepSeek-Coder-6.7B on SecurityEval), with significant implications for feedback mechanism selection. 
Models with low consistency introduce noise into feedback, making them unsuitable for iterative feedback loops regardless of average accuracy.
All models exhibit lower consistency on SecurityEval than SVEN. 
This reflects SecurityEval's hierarchical diversity (pillar, class, base, and variant weaknesses detailed in Section~\ref{subsec: dataset}) versus SVEN's uniform base-level labels.
The effect is particularly pronounced for smaller models.
Qwen2.5-Coder-7B drops from 71.64\% to 21.49\% Perfect Agreement between datasets. 
Model rankings remain highly stable across datasets (Spearman $\rho = 0.96$, $p<0.001$), indicating consistency is an intrinsic model property. 
Qwen2.5-Coder-32B and Devstral-24B consistently rank as most reliable, whilst DeepSeek-Coder-6.7B and Llama3.1-8B show highest variability. 
The gap between Perfect and Majority Agreement averages 24.1 percentage points on SVEN and 36.1 percentage points on SecurityEval, revealing that whilst models often reach two-run consensus, complete consistency remains challenging.

\section{Discussion}
ALPHA is, to our knowledge, the first Python-focused benchmark that empirically separates the performance characteristics of SAST tools and LLMs at the CWE level using a taxonomy-aware penalty. 
While LLMs achieve substantially stronger ALPHA scores, SAST tools exhibit a complementary profile: low coverage but relatively high precision when a detection is produced. 
Reporting both ALPHA and plain accuracy reveals this distinction and motivates a hybrid approach, where SAST outputs are prioritised when available and LLMs serve as a fallback or refinement mechanism within an iterative loop.

The substantial standard deviations observed, particularly for DeepSeek-Coder ($\pm 194.3$ on SVEN), reflect both genuine model stochasticity and the temperature=0.8 sampling setting. 
Consistency should thus be treated as a first-class criterion alongside raw performance.

Notably, all models exhibit lower consistency on SecurityEval than SVEN (Tables~\ref{tab: consistency_sven} and ~\ref{tab: consistency_sec_eval}).
The dataset characteristics detailed in Section~\ref{subsec: dataset} provide a concrete explanation
which validates ALPHA's design choice to penalise over-generalisation more heavily than over-specification when ground truth spans multiple abstraction levels, models must navigate varying specificity requirements.

Direct comparison of ALPHA scores between datasets is inadvisable due to differing sample sizes (121 vs 342) and difficulty distributions. 
However, model rankings remain stable (Spearman $\rho=0.86$), indicating that relative performance is dataset-independent even when absolute scores vary.

The temperature setting (0.8) introduces deliberate randomness reflecting real-world LLM deployment, where deterministic decoding (temperature=0) is often impractical due to reduced output quality. 
Our consistency analysis therefore measures variance under realistic conditions rather than artificially constrained settings.

The hierarchical structure of ALPHA naturally lends itself to use as a training signal. 
Unlike binary classification losses, ALPHA provides continuous, direction-aware penalties that could guide models toward hierarchy-aware representations. 
We envision augmenting supervised fine-tuning with ALPHA penalties through a dual-head architecture, though empirical validation of this approach remains essential future work (see Section~\ref{subsec: con3}).

The ALPHA penalty parameters $(\alpha_{up} = 2.0, \alpha_{lateral} = 1.8, \alpha_{min} = 1.1)$, are defined by construction to ensure the hierarchical ordering, 
$\alpha_{up} > \alpha_{lateral} > \alpha_{down}$ .
Alternative parameterisations may yield different absolute scores whilst preserving relative rankings.

To facilitate reproducibility, all code, data, and experimental configurations will be released upon publication.

\section{Future Work}
\label{subsec: con3}
A key design goal of ALPHA is its direct applicability as a training signal for model improvement. 
Unlike binary accuracy metrics, ALPHA provides continuous, direction-aware penalties that encode semantic relationships between CWE predictions. 
We propose augmenting supervised finetuning with ALPHA penalties to guide models towards hierarchy-aware representations.

We propose an extended transformer architecture with a classification head operating in parallel with the generative language modelling head.
Given an input sequence $x = (x_1, ..., x_n)$, the transformer produces hidden states $\mathbf{H} = [h_1, ..., h_n] \in \mathbb{R}^{n \times d_{hidden}}$ where $d_{hidden}$ is the hidden dimension of the LLM.
The classification head aggregates $\mathbf{H}$ via mean pooling $\bar h = \frac{1}{n} \sum^n_{i=1} h_i$ and projects to the CWE label space through either a single linear layer or a multi-layer perceptron (MLP): $z_{cls} = \mathbf{W}_{cls} \times \bar h + \mathbf{b}_{cls} \in \mathbb{R}^{|V|}$ where $\mathbf{W}_{cls} \text{ and } \mathbf{b}_{cls}$ are weight matrix and bias vector, $V$ denotes the set of all CWEs; yielding a probability distribution $p_\theta(v|x) = softmax(z_{cls})$ over CWE classes.

The training objective combines standard cross-entropy for text generation $\mathcal{L}_{CE}(\theta)$ with an expected penalty loss that directly incorporates ALPHA's hierarchical structure:
\begin{equation*}
    \mathcal{L}_{ALPHA}(\theta) = \sum_{v \in V} p_\theta(v|x) \cdot \hat{P}(v, c_{true})
\end{equation*}
where $\hat{P}(v, c_{true}) = \frac{P(v, c_{true})}{P_{max}}$ is the normalised ALPHA penalty with $P_{max} = d_{max} \times \alpha_{oog}$ and $d_{max}$ is the diameter of $G$.
The combined loss is - 
\begin{equation*}
    \mathcal{L}_{total} = \mathcal{L}_{CE}(\theta) + \lambda \cdot \mathcal{L}_{ALPHA}(\theta)
\end{equation*}
This formulation is fully differentiable, with gradients from both objectives flowing through the shared transformer backbone.
The expected penalty formulation weights each potential misclassification by both its probability and hierarchical severity, with ALPHA's direction-aware penalties naturally shaping the loss landscape.
Normalisation ensures $\hat P \in [0, 1]$, providing numerical stability and enables hyperparameter $\lambda$ tuning. 

The asymmetric penalty structure would theoretically induce:
\begin{enumerate}
    \item \textbf{Specificity preference}: Given uncertainty between a general CWE and its specific child, the model is incentivised to predict the more specific variant. 
    
    \item \textbf{Hierarchical coherence}: Predictions that remain within the correct branch of the CWE hierarchy receive lower penalties than those crossing to unrelated branches, encouraging semantically coherent errors.
    
    \item \textbf{Calibrated confidence}: The continuous penalty signal enables models to learn nuanced vulnerability relationships rather than treating all misclassifications equally.
\end{enumerate}
Future work will empirically validate this framework on larger datasets and measure downstream security improvements in deployed systems.

Additionally, future work will focus on deploying the best-performing LLMs and hybrid SAST–LLM systems in end-to-end iterative repair pipelines to directly measure remediation success. 
We will also investigate the use of the ALPHA penalty as a training loss, study prompt engineering and decoding constraints more systematically. 
Together, these directions aim to translate ALPHA’s evaluative insights into practical, deployable vulnerability remediation systems.
Our exclusive focus on open-source LLMs directly serves this objective. 
The ALPHA-guided training framework proposed in Section~\ref{subsec: con3} requires model weights to be accessible for fine-tuning—a capability exclusive to open-source models. 
By evaluating only models that can subsequently be trained using ALPHA penalties, we establish a complete pathway from benchmark evaluation to targeted improvement. 
Moreover, open-source models enable cost-effective deployment in iterative feedback loops without per-query API costs, ensuring our findings translate into economically viable, production-ready vulnerability detection systems.

\section{Limitations}

We explicitly constrained LLMs to produce a single CWE prediction to provide clear, actionable signals for iterative correction, critical for feedback-loop settings where multi-label outputs dilute corrective signals.
Although most models complied, a small subset produced multiple CWEs. 
Manual inspection indicated these cases reflected iterative refinement rather than independent detections; we selected the final reported CWE as the model's implicit best estimate.

This constraint creates an asymmetry with SAST tools, which inherently produce multiple predictions without configurability for single-output mode. 
Preliminary analysis of multi-label LLM outputs revealed substantially higher false positive rates, making them unsuitable for feedback systems. 
However, this asymmetry does not materially advantage SAST tools: the minimal difference between confidence-based selection and any-match strategies (Table~\ref{tab: alpha_results}) demonstrates that additional predictions rarely contain ground truth CWEs. 
Both approaches, therefore, evaluate tools under their most practical operating modes.

Our evaluation focuses on function-level Python vulnerability detection using two datasets (SecurityEval: 121 samples, SVEN: 342 samples). 
Findings may not generalise to other languages or granularities (e.g., repository-level). 

The dual-head architecture (Section~\ref{subsec: con3}) provides a theoretically sound pathway for incorporating ALPHA into supervised fine-tuning but lacks empirical validation. 
Practical effectiveness, including convergence behaviour, computational overhead, and downstream performance improvements, requires experimental assessment.

\section{Conculsion}
In this paper, we introduced ALPHA, a taxonomy-aware benchmark for evaluating CWE-level vulnerability identification in Python code. 
By leveraging the hierarchical structure of the CWE graph and a direction-sensitive penalty, ALPHA captures not only whether a vulnerability is detected, but also how close a prediction is to the ground truth in terms of actionable specificity. 
This enables a more faithful assessment of real-world utility than flat accuracy metrics.
Our empirical results demonstrate that large language models substantially outperform traditional SAST tools under ALPHA, particularly in their ability to identify fine-grained CWE variants. 
At the same time, SAST tools exhibit high precision when detections occur, highlighting a complementary trade-off that motivates hybrid system designs. 
We further show that constraining LLMs to produce single CWE predictions and analysing consistency across runs are essential considerations for feedback-driven remediation pipelines.
Beyond evaluation, ALPHA provides a principled foundation for future optimisation. 
The benchmark exposes clear signals that can inform loss and reward design for iterative vulnerability correction, while also revealing practical limitations such as model instability and taxonomy dependence. 
Overall, this work establishes ALPHA as both a diagnostic benchmark and a stepping stone towards robust, hybrid vulnerability remediation systems that integrate the strengths of LLMs and static analysis.

\bibliographystyle{IEEEtran}
\bibliography{references}

\appendix
\subsection{System Prompt}
\label{app: system_prompt}
\noindent You are a developer who is very security-aware with expertise in common vulnerability patterns.
For a given code block, perform a systematic security analysis.\\

\noindent CRITICAL CONSTRAINT: You MUST identify EXACTLY ONE CWE. 
Even if multiple vulnerabilities exist, select only the single most critical one.\\

\noindent INSTRUCTIONS:
1. Examine each line and function for potential security issues
2. Consider common attack vectors relevant to the code's apparent purpose
3. Assess the severity and exploitability of any findings
4. Identify the SINGLE most critical vulnerability present in the code
5. Report ONLY that one vulnerability using its CWE identifier\\

\noindent RESPONSE FORMAT (MANDATORY): \\
\#\#\# Analysis \\
$\langle$analysis$\rangle$
Detailed analysis of your findings, including reasoning for why this is the most critical vulnerability
$\langle$/analysis$\rangle$

\noindent \#\#\# Findings Summary\\
CWE: CWE-N \\

\noindent IMPORTANT NOTES:
- Use EXACTLY the format "CWE: CWE-N" where N is the number
- Do NOT list multiple CWEs (e.g., "CWE-89, CWE-20" is INVALID)
- If no vulnerability exists, write "CWE: None"
- In ``\#\#\# Findings Summary" section, you MUST not mention any other CWE other than the chosen CWE.\\

\noindent CORRECT EXAMPLE:\\
**Code Block:**\\
```python\\
def login(username, password):\\
\indent query = f``SELECT * FROM users WHERE \textbackslash \\
\indent username='{username}' AND password='{password}'"\\
\indent result = db.execute(query)\\
\indent if result:\\
\indent \indent return "Login successful"\\
\indent return "Login failed"\\
```

\noindent **Expected Response:**\\
\#\#\# Analysis\\
$\langle$analysis$\rangle$
This login function contains a critical SQL injection vulnerability. 
The function directly concatenates user input (username and password) into the SQL query string without any sanitisation or parameterisation. 
An attacker could manipulate the username parameter with malicious SQL code such as "admin' OR '1'='1' --" to bypass authentication entirely.
While the function also stores passwords in plaintext (CWE-256), the SQL injection (CWE-89) is more critical as it allows complete authentication bypass and potential database compromise. Therefore, CWE-89 is identified as the most critical vulnerability.
$\langle$/analysis$\rangle$

\noindent \#\#\# Findings Summary\\
CWE: CWE-89


\end{document}